\documentclass[aps,prl,twocolumn,showpacs]{revtex4}
\usepackage{amssymb}
\usepackage[dvips]{graphicx}
\usepackage{dcolumn}
\usepackage{bm}
\usepackage{amsmath}
\usepackage{amsfonts}
\input epsf.sty

\def\({\left(}
\def\){\right)}

\begin{document}

\title{Effects of spatial non-uniformity on laser dynamics.}
\author{L. I. Deych}
\affiliation{Physics Department, Queens College of City University
of New York, Flushing, NY 11367}

\begin{abstract}
Semiclassical equations of lasing dynamics are re-derived for a
lasing medium in a cavity with a spatially non-uniform dielectric
constant. The non-uniformity causes a radiative coupling between
modes of the empty cavity, which results in a renormalization of
self- and cross-saturation coefficients. Possible manifestations
of these effects in random lasers are discussed.
\end{abstract}

 \maketitle

\paragraph{Introduction}
Random lasers, in which optical feedback is provided by scattering
of light due to spatial inhomogeneity of the medium rather than by
well defined mirrors, has recently attracted a great deal of
attention\cite{CaoReview,VardenyRaikhIEEE2003}. In the case of
weak scattering, when the propagation of light can be described
within diffusion approximation, the nature of lasing in such
systems has been well understood starting with a pioneering work
by Letokhov\cite{Letokhov} followed by a large volume of
subsequent experimental and theoretical studies. The case of
strong scattering, however, when light can be at the verge of
Anderson localization, remains much more controversial.
Experimental results of Ref.\cite{CaoPRL1999} and consecutive
works with strongly scattering systems (see recent reviews in
Ref\cite{CaoReview} and Ref.\cite{VardenyRaikhIEEE2003}) led to an
assumption that lasing observed in those experiment is due to
formation of pre-localized, if not completely localized states of
light, which play a role of lasing cavities\cite{RaikhPRL2002},
and provide \emph{coherent resonant} optical feedback as oppose to
\emph{non-resonant feedback} affecting only intensity of light in
the diffusion case. The presence of narrow multiple lasing
peaks\cite{CaoPRL1999} as well as Poisson statistics of emitted
radiation\cite{CaoPRL2001} were considered as evidences in the
favor of this interpretation of these experiments. However, it was
shown in Ref.\cite{WiersmaPRL2004} that the non-resonant feedback
can also result in lasing with multiple narrow peaks. Moreover,
the authors of Ref.\cite{JohnPRL2004} demonstrated that the
Poisson statistics also cannot be considered as an exclusive
attribute of lasing with the resonant feedback.

In this situation, the recent results of
Ref.\onlinecite{CaoPRB2003} assume a particular significance. In
these experiments a multi-peak lasing was observed in $PMMA$
sheets containing a rhodamine dye as an active material and
titanium dioxide microparticles as scatterers. This system is
characterized by a strong inhomogeneous broadening of the lasing
transition, and most of the lasing peaks are separated in the
frequency domain by a homogeneous line width of the lasing
transition, $\gamma_a$. This is naturally explained by the
competition of modes "feeding" from the same population inversion
and spectral hole burning in inhomogeneously broadened
systems\cite{CaoPRB2003}. However, above certain value of the
pumping intensity, there were observed two lasing peaks coexisting
\emph{within the homogeneous line width, $\gamma_a$} and having
synchronized temporal behavior . This observation is indicative of
the genuine two-mode lasing, which can occur in regular cavity
lasers, when the mode competition is weakened by spatial hole
burning\cite{Siegman_book}. Such a behavior, however, cannot take
place in the the case of the non-resonant feedback, because, in
diffusive systems lasing only occur at the frequency of an atomic
transition\cite{Letokhov} (multiple peaks in
Ref.\cite{WiersmaPRL2004} are due to inhomogeneous broadening of
the transition used to generate emission and do not signify a
truly multi-mode behavior).

Thus, as of today, the results of Ref.\onlinecite{CaoPRB2003}
provide the most convincing evidence of the resonant feedback in
random lasers. It is important, therefore, to achieve a clear
understanding of the specifics of non-linear mode interaction in
such systems. However, since the experiments of
Ref.\onlinecite{CaoPRB2003} deal with just a single realization,
the randomness, by itself, is not important here. What is
important is the spatial inhomogeneity of the quasi-cavity,
supporting the modes of interest. The main objective of this
paper, therefore, is to study how this spatial inhomogeneity
affects lasing threshold and spatial hole burning. Our
consideration, however, is  not constrained by random lasers, and
can be applied to any type of lasers with spatially non-uniform
cavities. Currently, there is a tremendous interest to lasing in
the systems with a modulated dielectric constant, for instance,
photonic crystals. The result presented here are relevant for
these systems as well. Moreover, current technologies allow for
engineering structures with virtually arbitrary spatial profile of
the dielectric function. The results of this paper can be used to
manipulate properties of lasers by using spatial dependence  of
the cavity dielectric function as a new design parameter.

A general multi-mode theory of lasing in systems with an
arbitrarily inhomogeneous dielectric constant, $\epsilon({\bf
r})$, presented here is an extension of semi-classical Lamb
theory\cite{Siegman_book} for the media whose dielectric constant
is inhomogeneous   \emph{in the direction of propagation} of the
laser beam (inhomogeneity in the perpendicular directions results
in waveguiding effects, which are well studied in laser physics
(see, for instance,\cite{Yariv_book}). This inhomogeneity modifies
the orthonormalization condition for the eigen modes of the
cavity, making the standard inner product of the modes belonging
to different eigen frequencies different from zero. The main
effect resulting from this non-orthogonality is a new type of
linear coupling between normal modes of the empty cavity, which is
mediated by the polarization of the active medium.

The non-orthogonality of eigen modes due to the inhomogeneity of
$\epsilon({\bf r})$ should not be confused with non-orthogonality
of  Fox-Li modes of \emph{uniform} but \emph{leaky (open)}
cavities, which arises due to \emph{non-hermitian} nature of the
respective eigen value problem and does not result in any
additional linear coupling between the modes\cite{Siegman_book}.
The only consequence of the non-hermitian nature of such cavities
is the presence of an additional factor in the linear
susceptibility of the active medium, which was carefully studied
in the past and shown to be responsible for the excess noise in
unstable cavities\cite{SiegmanPRA1989a}.
\paragraph{Multi-mode laser equations for an inhomogeneous medium}
We consider an ideal cavity specified by an inhomogeneous
dielectric function $\epsilon({\bf r})$ and some boundary
conditions. The cavity is filled with an active medium
characterized by its polarization ${\bf P}({\bf r})$. Let us
assume that we know the full system of eigen modes, $f_k({\bf
r})$, and respective eigen frequencies $\omega_k$ of such a cavity
in the absence of the polarization. These modes can be used to
present electric field, $E$, and polarization $P$ in the form of
their linear combinations: $E=\sum_k E_k(t)f_k({\bf r})$,
$P=\sum_k P_k(t)f_k({\bf r})$,   where we assume that only
s-polarized modes couple to the active medium, and ignore the
vector nature of the field and the polarization. The
orthonormalization condition for these modes involves
inhomogeneous dielectric function $\epsilon({\bf
r})$\cite{GlauberPRA1991}:$\int \epsilon({\bf r})f^*_{k_1}({\bf
r})f_{k_2}({\bf r})d{\bf r}=\delta_{k_1k_2}$, which means that the
wave functions $f_k({\bf r})$ themselves are neither normalized
nor orthogonal. As a result, the dynamic equations for the
amplitudes, $E_k$, takes the following form
\begin{equation}\label{eq:mode equation}
    \ddot{E}_k(t)+(\omega_k-i\gamma_k)^2  E_k(t)=-4\pi\sum_{k_1}
    V_{kk_1}\ddot{P}_{k_1}(t),
\end{equation}
where we introduced cavity losses, characterized by
phenomenological parameters $\gamma_k$. The main peculiarity of
Eq.(\ref{eq:mode equation}) is the presence of the linear coupling
between different polarization amplitudes, $P_k$, characterized by
non-diagonal elements of the matrix
\begin{equation}\label{eq:Vkk1}
V_{kk_1}=\int f^*_{k_1}({\bf r})f_{k_2}({\bf r})d{\bf r}
\end{equation}
The presence of such a coupling is the main difference between
homogeneous and inhomogeneous cavities. The magnitude of coupling
parameters $V_{kk_1}$ depends on the spatial profile of the
dielectric constant, and can be tailored to enhance (or diminish)
the coupling effects.

Similar equation can be in principle derived for inhomogeneous
open cavities as well, where eigen modes $f_k$ should be replaced
by appropriate Fox-Li modes. The hermitian orthogonality in this
case is replaced by the bi-orthogonality, which involves adjoint
set of modes. For a general case of non-uniform open cavity this
condition was derived, for instance, in
Ref.\onlinecite{DutraPRA2000}. We shall leave, however, this topic
for future work.

A gain medium is described within the model of two-level atoms,
characterized by dephasing rate $\gamma_a^{-1}$, and population
relaxation time $\tau$, and we use a standard density matrix
approach in order to derive equations for polarization amplitudes
$P_k$, and population difference $\Delta N$. The next standard
step in the derivation of rate equations would be rotating wave
and slow amplitude approximations, which amount to presenting mode
amplitudes $E_k$ and $P_k$ as $E_k(t)=\Xi_k(t)\exp(-i\Omega_k t)$,
$P_k(t)=\Pi_k(t)\exp(-i\Omega_k t)$, where $\Xi_k$ and $\Pi_k$ are
slowly changing amplitudes, and $\Omega_k$ is a frequency of the
respective lasing mode. However, forcing this procedure onto
equation (\ref{eq:mode equation}) yields linear oscillating terms
of the form $\sum
V_{kk_1}\chi^{(1)}(\Omega_{k_1})\exp[-i(\Omega_k-\Omega_{k_1})]\Xi_{k_1}$
,which render derivation of meaningful rate equations impossible.
Here
\begin{equation}
\chi^{(1)}(\omega)=\frac{|\mu|^2\Delta
N_0}{4\hbar}\frac{1}{\omega-\omega_0+i\gamma_a}
\end{equation}
is a linear susceptibility of the gain medium with $|\mu|$  being
dipole matrix element of the lasing transition. Parameter $\Delta
N_0$ represents non-saturated population inversion, and
characterizes the strength of the pumping.

The physical origin of this problem is quite clear --- in the
presence of the linear coupling the modes of a passive medium are
not genuine normal modes of the entire system. As a result, an
attempt to excite such a mode leads to exchange of energy between
coupled modes and to non-stationary oscillations of the respective
intensities. The rate equations, therefore, should be derived for
the normal modes of the entire system, which would include cavity
and the gain medium. To this end, it is convenient to transform
Eq.(\ref{eq:mode equation}) in the frequency domain using
conventionally defined Fourier transformation:
\begin{eqnarray}\label{eq:fourier-transform}
&\sum_{k_1}\left[\left(\omega_k-i\gamma_k-\omega\right)\delta_{kk_1}-2\pi\omega_0V_{kk_1}\chi^{(1)}(\omega)\right]\tilde{E}_{k_1}(\omega)&\\
&=2\pi\omega_0\sum_{k_1}V_{kk_1}\tilde{P}^{(3)}_{k_1}(\omega)&\nonumber
\end{eqnarray}
where a tilde on the top of a symbol signifies the Fourier
transform of the respective quantity, and the polarization is
separated into a linear and third-order non-linear contribution,
${\tilde P^{(3)}}$. The former is taken into account in
Eq.(\ref{eq:fourier-transform}) by introducing a linear
susceptibility $\chi(\omega)$, and the expression for the latter
was derived in a standard way from the full system of density
matrix equation using standard perturbation approach.
\begin{eqnarray}\label{eq:non-linear_polariz}
\tilde{P}_k^{(3)}=\displaystyle{\frac{|\mu|^4\Delta
N_0}{32\pi^2\hbar^3}}\sum_{kk_1k_2k_3}A_{kk_1k_2k_3}\int
d\omega_1d\omega_2\times\\
\displaystyle{\frac{\tilde{E}_{k_1}(\omega-\omega_1)}{(\omega-\omega_0+i\gamma_a)(i\omega_1-1/\tau)}
\left[\frac{\tilde{E}_{k_2}(\omega_2)\tilde{E}_{k_3}(\omega_1-\omega_2)}{i(\omega_0-\omega_2)+\gamma_a}+c.c\right]}\nonumber
\end{eqnarray}
Anticipating the future use of the rotation wave approximation
applied to genuine normal modes of the system (see
Eq.(\ref{mode_expansion}) below) I substituted
$2\omega(\omega-\omega_k+i\gamma_k)$ instead of
$\omega^2-(\omega_k-i\gamma_k)^2$, neglected the non-resonant part
of the linear susceptibility, and replaced all frequencies
$\omega$ in non-resonant expressions with atomic frequency
$\omega_0$. The latter approximation is justified because we will
only consider the case where frequencies of all participating
modes lie within a homogeneous line width of the lasing
transition. Non-linear coupling parameters in
Eq.(\ref{eq:non-linear_polariz}), $A_{kk_1k_2k_3}$, are defined as
\begin{equation}
A_{kk_1k_2k_3}=\int{\epsilon({\bf r})f_{k}^{*}({\bf
r})f_{k_1}({\bf r})f_{k_2}({\bf r})f_{k_3}^*({\bf r})d^3r}
\end{equation}

\paragraph{Lasing threshold and non-linear dynamics of the intensities}
In order to illustrate the effects of spatial inhomogeneities on
lasing threshold we find eigen frequencies of linearized
Eq.(\ref{eq:fourier-transform}) in  a two-mode case. Imaginary
parts of these frequencies, $\Gamma_{1,2}$ are both positive below
lasing threshold. With increasing pumping, however, one of them,
$\Gamma_1$, for instance, first changes its sign, and this point
determines the lasing threshold. If we assume that
$\gamma_1\ll\gamma_2$,  a simple expression for the lasing
threshold can be derived:
\begin{equation}\label{eq:threshold}
\Delta N_{0_{tr}}=\frac{\Delta
N_{0_{tr}}^0}{V_{11}}\left[1-\left(\frac{V_{12}}{V_{11}}\right)^2\frac{\gamma_1}{\gamma_2}\right]
\end{equation}
where $N_{0_{tr}}^0$ is a threshold value of $\Delta N_0$ in a
system with a uniform dielectric constant. Two effects of the
non-uniformity appear in this expression. First, factor $V_{11}$,
which would be equal to unity for a uniform medium, affects the
threshold even in the absence of the linear coupling between the
modes. The value of this parameter depends on the spatial profile
of the dielectric function; with an appropriate choice of the
latter one can achieve a decrease in the lasing threshold. The
second effect reflected in Eq.(\ref{eq:threshold}) is due to the
coupling between the modes and results in further decrease of the
threshold.

In order to derive rate equations we have to diagonalize the
linear part of Eq.(\ref{eq:fourier-transform}). To this end, we
will, first, neglect the dispersion of the linear susceptibility.
This approximation is justified if we are only interested in
dynamics of intensities rather than lasing frequencies, and
because all the frequencies of interest lie within the width of
the atomic transition. After that we have to solve the eigen-value
problem for the remaining matrix, which is, however, essentially
non-hermitian.  Therefore, we have to find two adjoint sets of
vectors --- right ($\mid e_i\rangle$) and left ($\langle
\tilde{e}_j\mid$), which obey the bi-orthogonality condition
$\langle\tilde{e}_j\mid e_i\rangle=0$ when $i\neq j$. In order to
preserve standard expressions for intensities we shall normalize
our right eigen-vectors using condition $\langle e_i^*\mid
e_i\rangle=1$ (so called power normalization\cite{Siegman_book}).
As a result the product $\alpha_i= \langle \tilde{e}_i\mid
e_i\rangle\neq 1$. In order to  eliminate linearly coupled terms
from Eq.(\ref{eq:fourier-transform}) we present cavity mode
amplitudes, $\tilde{E}_k$, as a linear combination of the right
eigen vectors, $\mid e_i\rangle$,
\begin{equation}\label{mode_expansion}
\tilde{E}_k(\omega)=2\pi\sum
T_{ki}\left[Z_i(\omega)\delta(\omega-\Omega_i)+Z_i(\omega)\delta(\omega+\Omega_i)\right],
\end{equation}
where columns of matrix $T_{ki}$ are formed by the vectors $\mid
e_i\rangle$.  In Eq.(\ref{mode_expansion}) we also introduced a
slow changing amplitude approximation applied to the amplitudes of
the true normal modes of the system. In the frequency domain, this
approximation consists of presenting the amplitudes as a product
of a frequency dependent part and a delta function, containing a
lasing frequency $\Omega_i$. Matrix $T_{ki}$ is not a unitary
matrix, and, transformation of Eq.(\ref{eq:fourier-transform}) to
the new basis has to rely on matrix $\tilde{T}_{ki}$, whose rows
consist of vectors $\langle \tilde{e}_i\mid$. The product of
matrices ${\bf T}$ and ${\bf \tilde{T}}$ is a diagonal matrix with
elements $\alpha_i$. The resulting equations for slow amplitudes
$Z_i$ take the form
\begin{equation}\label{decoupled_equations}
-i\frac{dZ_i}{dt}+\left(\tilde{\Omega}_i-\Omega_i\right)Z_i=2\pi\omega_0\sum_ju_{ij}\Pi_j^{(3)}e^{-i(\Omega_j-\Omega_i)t},
\end{equation}
where
$u_{ij}=\sum_{k,k_1}\tilde{T}_{ik}V_{kk_1}T_{k_1j}/\alpha_i$, and
the non-linear contribution to polarization, in the new basis, is
given by
\begin{equation}\label{nonlinear_polar}
\Pi_i^{(3)}=\frac{|\mu|^4\Delta
N_0\tau}{8\hbar^3\alpha_i\gamma_a}\sum_{j,l,m}R_{ijlm}\frac{Z_jZ_l^*Z_me^{-i(\Omega_j-\Omega_l+\Omega_m-\Omega_i)t}}{\Omega_j-\Omega_l+\Omega_m-\omega_0+i\gamma_a}
\end{equation}
\begin{equation}\label{nonlinear_coeff}
R_{ijlm}=-\sum_{{k_i}}\tilde{T}_{ik}A_{kk_1k_2k_3}T_{k_1m}\left[T_{k_2j}\left(T_{k_3l}\right)^*+T_{k_3j}\left(T_{k_2l}\right)^*\right]
\end{equation}
where  we have neglected frequency dependence of the nonlinear
coefficients $R_{ijlm}$. Equations (\ref{decoupled_equations}),
(\ref{nonlinear_polar}), and (\ref{nonlinear_coeff}) provide a
basis for further analysis of the non-linear dynamics of the
system under consideration.

In particular, the rate equations can be obtained in a standard
way by separating real and imaginary parts of
Eq.(\ref{decoupled_equations}) and neglecting all oscillating
terms on its right-hand side:
\begin{equation}
\frac{dI_i}{dt}=2I_i\left[-\Gamma_i-\beta_iI_i-\sum_j\theta_{ij}I_j\right]
\end{equation}
Here $I_i$ is the dimensionless intensity of the $i$-th mode,
$\Gamma_i$ is its unsaturated amplification rate, $\beta_i$ and
$\theta_{ij}$ are self- and cross-saturation parameters
respectively, which are expressed in terms of coefficients
$R_{ijlm}$ of Eq.(\ref{nonlinear_coeff}). These equations have the
same form as standard lasing rate equations for a uniform medium,
with the only difference being that instead of the combination of
non-saturated gain and loss terms, we have a single parameter
$\Gamma_i$, representing the imaginary part of the mode's eigen
frequency. The main effect of the linear coupling is a
renormalization of the non-linear coupling coefficients. The most
important feature of this renormalization, which makes it
experimentally relevant, is a non-trivial dependence of the new
coefficients on the intensity of pumping. This dependence arises
because the coupling is carried by polarization, and, hence, its
strength is proportional to the unsaturated inversion $\Delta
N_0$. In order to illustrate the last point we consider an example
of only two interacting modes. There are two possible regimes of
behavior in this case: single mode lasing, when the mode
competition prevents the second mode from lasing, and two-mode
lasing, when the spatial hole burning prevails over the
competition. The choice between these regimes is determined by a
coupling parameter $C$, defined by the ratio of cross- and
self-saturation parameters
$C=(\theta_{12}\theta_{21})/(\beta_{11}\beta_{22})$\cite{Siegman_book}.
In a uniform medium, this parameter depends solely upon spatial
distribution of the cavity modes, determined by the cavity's
geometry. In the situation considered, here, this parameter
becomes dependent on the pumping intensity. In order to illustrate
the possible character of such dependence, we simulated parameter
$C$ for a cavity which consists of two dielectric materials with
different dielectric constants $\epsilon_1$, and $\epsilon_2$.
\begin{figure}[tbp]
\includegraphics[width=3.5in]{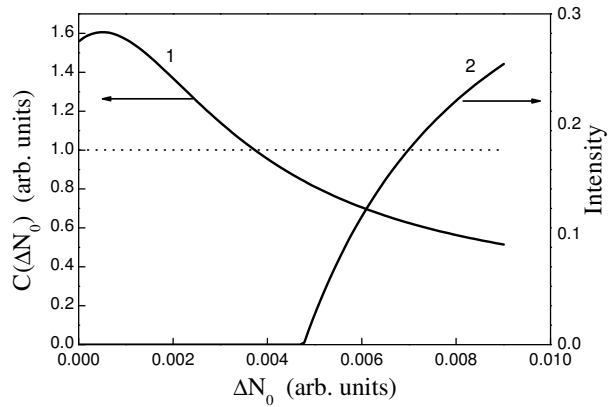}
\caption{Dependence of the non-linear coupling parameter $C$ on
$\Delta N_0$ (curve 1) and deviation of the intensity in a single
mode regime from a linear dependence on pumping (curve 2)}
\label{fig2}
\end{figure}
The curve 1 on the figure shows the dependence $C(\Delta N_0)$ for
two closest in frequency modes of such a cavity. The most striking
feature of this graph is the steep decrease of this coefficient
with $\Delta N_0$, which means that even if the modes of the empty
cavity wouldn't favor the spatial hole burning, the increasing
with pumping linear interaction between the modes modifies their
spatial structure in a way which is beneficial for the two-mode
lasing. Similar behavior of $C(\Delta N_0)$ was also found for the
dielectric constant of the shape $\epsilon(z)=\epsilon_0+az^2$ or
$\epsilon(z)=\epsilon_0+\delta\cos z$, where $z$ is coordinate in
the beam propagation direction. This effect might explain the two-
mode behavior observed in random lasers\cite{CaoSokoulisPRB2002}.
The fact that increased pumping can systematically drive $C$ below
unity for various configurations of $\epsilon({\bf r})$ makes such
effect much more likely to occur in a random system than just a
coincidental combination of various parameters suggested, for
instance in Ref.\cite{Soukoulis_coupling2004}. A presence of the
linear mode coupling in random lasers can be verified directly by
observation of dependence of $I(\Delta N_0)$ in single and
multi-mode regimes, which, if effects considered here are
responsible for the observed multi-mode behavior, should deviate
from a simple linear behavior expected in lasers with the uniform
dielectric constant (see curve 2 in Fig.1).

\paragraph{Conclusion}
We derived non-linear equations describing dynamics of lasing
modes in a cavity whose dielectric constant is spatially
non-uniform in the direction of beam propagation. For a number of
spatial profiles of $\epsilon({\bf r})$ it is shown that the
non-uniformity enhances spatial hole burning and promote two-mode
lasing. This effect can explain recent observation of two-mode
behavior in random lasers. The equations derived in the paper can
also be used to manipulate properties of lasers through a design
of spatial profile of the dielectric function.
\begin{acknowledgments}
This work is supported by AFOSR grant F49620-02-1-0305. Author
also would like to express his gratitude for stimulating
discussions with A. Lisyansky, H. Cao, R. Suris, and F. Capasso.
\end{acknowledgments}

\end{document}